\documentclass[12pt]{article}

\usepackage{amssymb}
\usepackage{amsmath}
\usepackage{latexsym}
\usepackage{yfonts}

\catcode`@=11
\renewcommand{\section}{\@startsection{section}{1}{0pt}{\medskipamount}
{\medskipamount}{\large\bf}} \numberwithin{equation}{section}
\catcode`@=12
\def\beq{\begin{eqnarray}}    %%%  begequation/eqnarray
\def\eeq{\end{eqnarray}}      %%%  endequation/eqnarray

\def\pa{\partial}                       %%% partial
\def\={\ =\ }

\begin{document}

\begin{center}
{\Large\bf Quantum antibrackets: polarization and parametrization}

\vspace{18mm}

{\Large Igor A. Batalin$^{(a,b)}\footnote{E-mail:
batalin@lpi.ru}$\;,
Peter M. Lavrov$^{(b, c)}\footnote{E-mail:
lavrov@tspu.edu.ru}$\;
}

\vspace{8mm}

\noindent ${{}^{(a)}}$
{\em P.N. Lebedev Physical Institute,\\
Leninsky Prospect \ 53, 119 991 Moscow, Russia}

\noindent  ${{}^{(b)}}
${\em
Tomsk State Pedagogical University,\\
Kievskaya St.\ 60, 634061 Tomsk, Russia}

\noindent  ${{}^{(c)}}
${\em
National Research Tomsk State  University,\\
Lenin Av.\ 36, 634050 Tomsk, Russia}

\vspace{20mm}

\begin{abstract}
\noindent By proceeding from a simple non-polarized formalism, we consider in detail
the polarization procedure as applied to the generating equations of the
quantum antibracket algebra, in terms of the parametrized generating operator.
\end{abstract}

\end{center}

\vfill

\\

\newpage

\section{Introduction}

Given a fermionic nilpotent operator $Q$, one can define, for any two operators
$X$ and $Y$, their quantum 2-antibracket \cite{BM,BM1},
\beq
\label{i1}
( X, Y )_{Q}  =:  \frac{1}{2}  ( [ X, [ Q, Y ] ]  -
 [ Y, [ Q, X ] ]  (-1)^{ ( \varepsilon_{X} + 1 ) ( \varepsilon_{Y} + 1 ) } ),  %      (0.1)
 \eeq
or the derived bracket in mathematical terminology \cite{K-S,Vo,CSch}. These objects have
very nice algebraic properties such as the anti-symmetry, the Jacobi relations,
and the Leibnitz rule. It  appears remarkably that the quantum 2-antibracket (\ref{i1})
arises naturally when studying BRST - invariant constraint algebra \cite{BT,BB}
as well as in formulating quantum dynamical equations \cite{BL} and
in the general representation of the gauge fields \cite{BL1}.

The quantum 2-antibracket (\ref{i1}) generalizes the usual
antibracket of the field-antifield BV formalism \cite{BV,BV1} (see
also \cite{SZ,ND,GST}). Indeed, when being operators $X$ and $Y$ in
(\ref{i1}) functions of the field-antifield variables only, one
identifies $Q  =  - \Delta$, where $\Delta$ is the odd Laplacian,
and then uses $[ \Delta, Y ]  =  ( \Delta Y )  + {\rm ad}( Y )
(-1)^{ \varepsilon_{Y} }$, to get $( X, Y )_{Q}=  ( X, Y )^{ (BV)
}$.

 In the present paper, we proceed from the non-polarized version of the (\ref{i1})
as defined at $X = Y = B$, with any bosonic operator $B$,
 \beq
\label{i2}
( B, B ) _{Q}  =: -  [ B, [ B, Q ] ].     %    (0.2)
 \eeq
So, we call the (\ref{i1}) as the polarized quantum 2-antibracket. In fact, that is
just being $X \neq Y$ we mean as the polarization, in the general sense. Given the
(\ref{i2}), one can polarize that by considering formally $B = \alpha X + \beta Y$, with
$\alpha$ and $\beta$ being the respective parameters,
and then taking the $\alpha\beta$ derivative of the (\ref{i2}),
 \beq
\label{i3}
( X, Y )_{Q}  =  \partial_{\alpha} \partial_{\beta}
 \frac{1}{2} ( B, B )_{Q} (-1)^{ \varepsilon_{Y} }.     %  (0.3)
 \eeq
It appears that the basic definitions and the general analysis look simpler
essentially in the non-polarized formalism.
     On the other hand, the generating equations of the quantum antibracket
algebra can be formulated naturally in terms of the parametrized generating
operator,
 \beq
\label{i4}
\tilde{Q}  =:  \exp\{ \lambda^{a} f_{a} \} Q \exp\{ - \lambda^{a} f_{a} \},   %   (0.4)
 \eeq
being $\{ f_{a} \} $ a chain of operators, and being $\{ \lambda^{\alpha} \}$
the respective
parameters. In this way, the generating equations of the quantum antibracket
algebra do acquire their geometrically-covariant status. We call the representation
(\ref{i4})  as the parametrization. All higher quantum antibrackets are defined in terms
of derivatives of the generating operator (\ref{i4}) with respect to the parameters.
Complete set of the structure relations of the quantum antibracket algebra is
generated by the nilpotence  of the operator (\ref{i4}).

\section{Basics on quantum antibrackets}

Let $Q$ be a fermionic nilpotent operator,
 \beq
\label{r1}
Q:    \varepsilon(Q)  =  1,  \quad Q^{2}  =  0,
\eeq
and let $B$ be an arbitrary bosonic operator,
 \beq
\label{r2}
B:     \varepsilon(B)  =  0.
 \eeq
Introduce a quantum 2-antibracket,
 \beq
\label{r3}
( B, B )_{Q}  =:  -  [ B, [ B, Q ] ],
 \eeq
then we have the main property
 \beq
\label{r4}
[  Q, ( B, B )_{Q} ]  =  [ [ Q, B ], [ Q, B ] ].
 \eeq
Let $A$ be the associator multiplied with a parameter $\beta$,
 \beq
\label{r5}
A  =:  \beta ( B, ( B, B )_{Q} )_{Q}.
 \eeq
It follows then
\beq
\label{r6}
[ Q, A ]  =  \beta [ [ Q, B ], [ [ Q, B ], [ Q, B ] ] ]  =  0 \;
 \Longrightarrow  \;A  =  [\mathcal{A}, Q].
 \eeq
By choosing the operator  $\mathcal{A}$ as
 \beq
\label{r7}
\mathcal{A}  =  [ B, (B, B)_{Q} ]  =  ( B, B, B )_{Q}  =:  -  [ B, [ B, [ B, Q ] ] ]  \;
 \Longrightarrow\; \beta  =  6,
 \eeq
we arrive at the non - polarized form of the Jacobi relation
 \beq
\label{r8}
6 ( B, ( B, B )_{Q} )_{Q}  =  [ ( B, B, B )_{Q}, Q ],
 \eeq
as an identity with respect to $B$.
 Indeed, denote the operators
\beq
\label{r9}
X  =:  [ B, [ [ B, Q ], [ B, Q ] ] ],   \quad
Y  =:  [ [ B, Q ], [ B, [ Q, B ] ] ],
 \eeq
such that
\beq
\label{r10}
X  =  - 2 Y.
\eeq
Then, we have
 \beq
\label{r11}
( B, ( B, B )_{Q} )_{Q}  =  \frac{1}{2} ( X + Y )  =  - \frac{1}{2} Y,
\eeq
 \beq
\label{r12}
[ ( B, B, B )_{Q}, Q ]  =  X - Y  =  - 3 Y,
 \eeq
which is equivalent to (\ref{r8}).

\section{Polarization}

In the definition (\ref{r3}), consider the Boson $B$ of the form
\beq
\label{p1}
B = \alpha X  +  \beta Y  +  \gamma Z,
\eeq
with $\alpha, \beta $ and $\gamma$ being parameters. It follows then the polarized
quantum 2-antibracket
 \beq
\label{p2}
\partial_{\alpha} \partial_{\beta} \frac{1}{2} ( B, B )_{Q} (-1)^{ \varepsilon_{Y} }
 = ( X, Y )_{Q}  =:  \frac{1}{2}  (  [ X, [ Q, Y ] ]  -
 [ Y, [ Q, X ] ]  (-1)^{ ( \varepsilon_{X} + 1 ) ( \varepsilon_{Y} + 1 ) }  ).
 \eeq
In analogy with the main property (\ref{r4}) we have its counterpart
for polarized quantum \\ 2-antibracket (\ref{p2})
\beq
\label{p2a}
[Q,(X,Y)_Q]=[[Q,X],[Q,Y]].
\eeq
In turn, by using (\ref{p2}), we have the polarized version of the Jacobi relation (\ref{r8})
\beq
\nonumber
&&( X, ( Y, Z )_{Q} )_{Q}  (-1)^{ ( \varepsilon_{X} + 1 ) ( \varepsilon_{Z} + 1 ) }  +
 {\rm cyclic \;permutations} ( X, Y, Z )  = \\
\nonumber
&&= \partial_{\alpha } \partial_{\beta} \partial_{\gamma}
 ( B, \frac{1}{2} ( B, B )_{Q} )_{Q}  (-1)^{ ( \varepsilon_{X} + 1 )
 ( \varepsilon_{Z} + 1 ) + \varepsilon_{Y} }  =\\
\nonumber
&&=
 \partial_{\alpha} \partial_{\beta} \partial_{\gamma} \;\!\frac{1}{2} \!\;
 \frac{1}{6} \;[ ( B, B, B )_{Q}, Q ]
 (-1)^{ ( \varepsilon_{X} + 1 ) ( \varepsilon_{Z} + 1 ) + \varepsilon_{Y} }  =\\
&&=
 \frac{1}{2} [ ( X, Y, Z )_{Q}, Q ]
 (-1)^{ ( \varepsilon_{X} + 1 ) ( \varepsilon_{Z} + 1 ) }.
\label{p3}
 \eeq
Thus, we identify the polarized quantum 3-antibracket,
 \beq
\label{p4}
&&( X, Y, Z )_{Q}  =: \partial_{\alpha} \partial_{\beta} \partial_{\gamma}
 \frac{1}{6}  ( B, B, B )_{Q}  (-1)^{ \varepsilon_{Y} }= \\
\nonumber
&&=
-  (-1)^{ ( \varepsilon_{X} + 1 ) ( \varepsilon_{Z} + 1 ) }  \frac{1}{3}
 \big(  [ X, ( Y, Z )_{Q} ]  (-1)^{ [ \varepsilon_{X} ( \varepsilon_{Z} + 1 ) +
 + \varepsilon_{Y} ] }  +  {\rm cyclic \;permutations}\; ( X, Y, Z ) \big ).
 \eeq
The modified Leibnitz rule for quantum 2-antibracket (\ref{p2})  reads
\beq
\nonumber
&&( XY, Z )_{Q}  -  X ( Y, Z )_{Q}  - ( X, Z )_{Q} Y (-1)^{ \varepsilon_{Y} (
\varepsilon_{Z} + 1 ) } = \\
\label{j217}
&&=\frac{1}{2} \big( [ X, Z] [ Y, Q ] (-1)^{ \varepsilon_{Z} ( \varepsilon_{Y} + 1 ) }
+  [ X, Q ] [ Y, Z ] (-1)^{ \varepsilon_{Y} } \big).
\eeq

\section{Parametrization}

Here we include in short the generating equations for the quantum
antibracket algebra \cite{BM1}.  Let us introduce an operator valued exponential
\beq
\label{A1}
U = \exp\{ \lambda^{a} f_{a} \},    \quad   U|_{\lambda = 0} = 1,
\eeq
 where $\{ f_{a},  a = 1, 2, ...  \}$,  is a chain of operators,  $\varepsilon(
f_{a} ) = \varepsilon_{a}$,  and $\lambda^{a}$ are parameters, $\varepsilon(
\lambda^{a} )  =  \varepsilon_{a}$.  Introduce the
$U$-transformed $Q$-operator,
\beq
\label{A2}
\tilde{Q}  =  U Q U^{-1},   \quad  \tilde{Q}^{2} = 0.
\eeq
The latter equation (\ref{A2}) does  generate the complete set of the higher
Jacobi relations  following  (\ref{p4}).

Further,  we have the generating equations
\beq
\label{A3}
\pa_{a} \tilde{Q}  =  [ R_{a}, \tilde{Q} ],   \quad   R_{a} =  ( \pa_{a} U )
U^{-1},    \quad     \pa_{a} = \frac{\pa}{\pa\lambda^{a}}, \quad
     \tilde{Q}|_{\lambda = 0} = Q,
\eeq
 \beq
\label{A4}
\pa_{a} R_{b}  -  \pa_{b} R_{a} (-1)^{ \varepsilon_{a} \varepsilon_{b} }  =  [
R_{a}, R_{b}] .
\eeq
It follows from (\ref{A4}) that  the  equation holds with the Euler  operator,
$N=: \lambda^{a} \partial_{a} $,
 \beq
\label{A4a}
( N  +  1 )  R_{b}  =  \partial_{b} \Psi  -  [ R_{b}, \Psi ],
\eeq
where
 \beq
\label{A4b}
\Psi  =  \lambda^{a} R_{a},
 \eeq
is an operator describing the arbitrariness in a choice of  $\lambda$
parametrization.

By making the rescaling
 \beq
\label{A4c}
&&\lambda^{a}\; \rightarrow \; t \lambda^{a},   \\
 \label{A4d}
&&
R_{a}  \; \rightarrow \;  \tilde{ R }_{a}   =:  t R_{a}( t \lambda ),   \\
 \label{A4e}
&&
\Psi \; \rightarrow \;  \tilde{ \Psi }  =:  \lambda^{a}R_{a}(t \lambda),
 \eeq
we convert the equation (\ref{A4a}) to the form
\beq
\label{A4f}
\frac{\partial \tilde{ R }_{b}}{\partial t}  =  \partial_{b} \tilde{ \Psi }  -
 [ \tilde{ R }_{b}, \tilde{ \Psi } ],
 \eeq
with the boundary condition
 \beq
\label{A4g}
\tilde{ R }_{b} |_{ t = 0 }  =  0.
 \eeq
The Cauchy problem (\ref{A4f}), (\ref{A4g}) resolves in the form
 \beq
\label{A4h}
\tilde{ R }_{b}( t )  =  \int_{0}^{t}  dt'  \;\!U( t, t' )
( \partial_{b} \tilde{ \Psi }( t' ) )  U^{-1}( t, t' ),
 \eeq
where $U( t, t' )$ resolves the Cauchy problem
 \beq
\label{A4j}
\frac{\partial U( t, t' )}{\partial t}  =  \tilde{ \Psi }( t )  U( t, t' ),
\eeq
\beq
 \label{A4k}
U( t, t' ) |_{ t = t' }  =  1.
\eeq

In parallel to the above geometric formulae (\ref{A4f}) - (\ref{A4k}), we suggest a
simpler derivation for the "current" $R_{a}$ (\ref{A3}),
\beq
\nonumber
R_{a}  &=&  \int_{0}^{1}  dt\;\!  \exp\{ t \psi \}  \;\!( \partial_{a} \psi  )\;\!
\exp\{ - t \psi \}  =
\int_{0}^{1}  dt \;\! \exp\{ t \;\!{\rm ad}( \psi ) \} \;\! ( \partial_{a} \psi  )  =\\
\label{A4l}
&=&\frac{ \exp\{ {\rm ad}( \psi ) \} - 1 }{ {\rm ad}( \psi ) }\;
( \partial_{a} \psi ),
\eeq
where
\beq
\label{A4z}
\psi=\lambda^a f_a,
\eeq
and we have used the identity
\beq
\label{A4z1}
[A, \exp\{B\}]=\int_0^1 dt \exp\{t B\}[A,B]\exp\{(1-t)B\},
\eeq
for $A=\partial_a$, $B=\psi$.

Notice that the following relation holds between the $\Psi$, (\ref{A4b}),
and the $\psi$, (\ref{A4z}),
\beq
\label{A4v}
\Psi=\frac{ \exp\{ {\rm ad}( \psi ) \} - 1 }{ {\rm ad}( \psi ) }\;
( N \psi ).
\eeq

By multiplying the equation (\ref{A4f}) with $\lambda^{b}$ from the left we get an identity,
as expected, which implies that the $\tilde{\Psi}$, (\ref{A4e}), is an arbitrary operator.
Then,
by choosing in (\ref{A4h}) - (\ref{A4k}) $\tilde{\Psi} = \psi$, with $\psi$ being given in
(\ref{A4z}), we arrive
at the representation (\ref{A4l}). Indeed, it follows from (\ref{A4f})
that the equation
\beq
\label{A4t}
\partial_{t} \;\!t \;\!\tilde{ \Psi }  =  N \tilde{ \Psi }
\eeq
holds, which implies in turn
\beq
\label{A4u}
t \;\!\partial_{t}\;\! \tilde{ \Psi }  =  \lambda^{a} N R_{a}( t \lambda ).
\eeq
That is just an identity expected. Due to (\ref{A4e}), the  left-hand side of
(\ref{A4u}) rewrites in the form
\beq
\label{A4o}
\lambda^{a} t \lambda^{b} \frac{ \partial }{ \partial (t \lambda^{b}) }\;\!
R_{a}( t \lambda )  =
\lambda^{a} N R_{a}( t \lambda ).
\eeq
Thus, the $\tilde{\Psi}$, (\ref{A4e}), remains arbitrary.

The Lie equation (\ref{A3}) and the  Maurer-Cartan equation
(\ref{A4}) do serve as the generating equations for quantum
antibrackets.  Here we present explicitly only the case of quantum
2-antibracket.   It follows from (\ref{A3}) by  $\lambda$
differentiating, that
\beq
\nonumber &&- \pa_{a} \pa_{b} \tilde{Q}
(-1)^{\varepsilon_{b}}  + \frac{1}{2} [ ( \pa_{a} R_{b} + \pa_{b}
R_{a} (-1)^{ \varepsilon_{a}
\varepsilon_{b} } ) (-1)^{\varepsilon_{b}} ,  \tilde{Q}  ]  =  \\
\label{A5}
&&=\frac{1}{2} \left(  [ R_{a}, [ \tilde{Q}, R_{b} ] ]  -
( a \leftrightarrow b ) (-1)^{ (
\varepsilon_{a} + 1 ) ( \varepsilon_{b} + 1 ) }  \right) =  ( R_{a}, R_{b} ) _{
\tilde{Q} }.
\eeq
In turn, it follows from (\ref{p4}) and (\ref{A5}) that the next equation holds,
\beq
\nonumber
&&-  \frac{1}{3}  (-1)^{ ( \varepsilon_{a} + 1 ) ( \varepsilon_{c} +
1 ) }
\big( [ R_{a}, \Delta_{bc} \tilde{Q} ]  (-1)^{ ( \varepsilon_{a} (
\varepsilon_{c} + 1 ) + \varepsilon_{b}) }  +\\
\label{A5a}
&&\qquad +\;
{\rm cyclic\; permutations}\; ( a, b, c ) \big)  =
( R_{a}, R_{b}, R_{c} )_{\tilde{Q} },
\eeq
where we have denoted
\beq
\label{A5b}
\Delta_{ab}  =:  - \partial_{a} \partial_{b}  (-1)^{ \varepsilon_{b}
}  +  [ Y_{ab}, ( \cdot ) ],
\eeq
\beq
\label{A5c}
Y_{ab}  =:  \frac{1}{2}  \big(  \partial_{a} R_{b}  +
\partial_{b} R_{a}  (-1)^{ \varepsilon_{a} \varepsilon_{b} }  \big)
(-1)^{ \varepsilon_{b} }.
\eeq
In the latter notation, the equation (\ref{A5}) takes the form
\beq
\label{A5d}
\Delta_{ab} \tilde{Q}  =  ( R_{a}, R_{b} )_{ \tilde{Q} }.
\eeq

It follows from (\ref{A5}) at $\lambda = 0$,
\beq
\label{A6}
-  ( \pa_{a}\pa_{b} \tilde {Q} ) (-1)^{ \varepsilon_{b} } |_{\lambda = 0}  =
( f_{a}, f_{b} )_{Q},
\eeq
 where we have used
\beq
\label{A7}
 ( \pa_{a} R_{b} ) |_{\lambda = 0}  =  \frac{1}{2}  [ f_{a}, f_{b} ].
\eeq
The next equation (\ref{A5a}) takes the form
\beq
\label{A7z}
\Delta_{abc} \tilde{Q}  =  ( R_{a}, R_{b}, R_{c} )_{ \tilde{Q} },
\eeq
where
\beq
\label{A7x}
\Delta_{abc}  =  -  \partial_{a} \partial_{b} \partial_{c}  (-1)^{ \varepsilon_{b} }  +
\;{\rm expression\; vanishing \;at}\; \lambda  =  0.
\eeq
In its more explicit form, the second  term in right-hand side in
(\ref{A7x}) reads
\beq
\label{A7y}
\nonumber
{\rm expression}  &=:&  \frac{1}{3}
(-1)^{ ( \varepsilon_{a} + 1 ) (\varepsilon_{c } + 1 ) }
\big\{  \big(  \partial_{b}  [ \partial_{c} R_{a}, ( \cdot ) ]  (-1)^{
\varepsilon_{a} ( \varepsilon_{b} +
 \varepsilon_{c} ) }  +  [ \partial_{b} R_{a}, \partial_{c}
( \cdot ) ]  (-1)^{ \varepsilon_{a}\varepsilon_{b} }  +\\
\nonumber
&&+\;
[ R_{a},  [ \frac{1}{2} (  \partial_{b} R_{c}
+  \partial_{c} R_{b}
(-1)^{ \varepsilon _{b} \varepsilon_{c} }  ), ( \cdot )  ]  ]
\big)(-1)^{ ( \varepsilon_{a} + 1 )( \varepsilon_{c} + 1 ) + \varepsilon_{b} }  +\\
\label{A7y}
&&
+ \;{\rm cyclic\; permutations}\; (a, b, c )  \big \}.
\eeq

It follows in a similar way that higher $\lambda$  derivatives of $\tilde{Q}$ do
yield all higher quantum antibrackets,
\beq
-  ( \pa_{a_1}\cdots\pa_{a_n} \tilde {Q} ) (-1)^{ E_n} |_{\lambda = 0}  =
( f_{ a_{1} } ,  ...  , f_{ a_{n} } )_{Q}   =  - {\rm Sym}( [ f_{
a_{1} }, ...  , [ f_{ a_{n} }, Q ] ... ] )  (-1)^{ E_{n} },
\eeq
where  we have denoted
\beq
E_{n} =  \sum_{ k = 1 }^{ [ n/2 ] } \varepsilon_{ a_{2k} },
\eeq
\beq
{\rm Sym}( X_{ a_{1}  ...  a_{n} } )  =  S_{ a_{1}  ...  a_{n} }^{ b_{n}
... b_{1} }  X_{ b_{1}  ...  b_{n} },
%\eeq
%\beq
\quad n!  S_{ a_{1}  ...  a_{n} }^{ b_{n}  ...  b_{1} }  =  \pa_{ a_{1} }
... \pa_{ a_{n} }  \lambda^{ b_{n} }  ...  \lambda^{ b_{1} }.
\eeq
It has also been shown in \cite{BM1,Ber}, how these equations enable one
to derive the modified Jacobi relations for subsequent higher
quantum antibrackets.

\section{Summary}

In the present article  we have considered a simple non-polarized
form of the quantum antibracket algebra (Section 2), and then
derived its polarized form (Section 3). Also, we have introduced
a natural parametrization (\ref{A2}), and then derived the respective
generating equations (\ref{A5d}), (\ref{A7z}) (Section 4). In an obvious way
the construction can be extended to cover all higher quantum
antibrackets.

\section*{Acknowledgments}
\noindent The authors would like to thank Klaus Bering for useful
discussions. The work of I. A. Batalin is supported in part by the
RFBR grant 17-02-00317. The work of P. M. Lavrov is supported by the
Ministry of Education and Science of Russian Federation, grant
3.1386.2017 and by the RFBR grant 18-02-00153.

\end{document}